\documentclass[reprint,superscriptaddress,amssymb,amsmath,aps,showpacs,10pt,longbibliography,prl]{revtex4-2}
\usepackage{graphicx}
\usepackage{enumitem}

\def\cm{cm$^{-1}$}
\newcommand{\parl}{\,\|\,}
\newcommand{\per}{\,\perp\!}

\graphicspath{{figures/}}
\usepackage{color}
\usepackage[colorlinks,bookmarks=false,citecolor=darkblue,linkcolor=red,urlcolor=blue]{hyperref}

\definecolor{darkred}{rgb}{0.7,0.0,0.0}
\definecolor{darkblue}{rgb}{0,0.02,0.45}

\def\cdbl{\color{darkblue}}
\definecolor{darkgreen}{rgb}{0.02,0.45,0.0}

\begin{document}


\title{Fermi-liquid behavior of non-altermagnetic RuO$_2$}

\author{Maxim Wenzel}
\affiliation{1. Physikalisches Institut, Universit{\"a}t Stuttgart, 70569 Stuttgart, Germany}

\author{Ece Uykur}
\email{e.uykur@hzdr.de}
\affiliation{Helmholtz-Zentrum Dresden-Rossendorf, Inst Ion Beam Phys \& Mat Res, D-01328 Dresden, Germany}

\author{Sahana R\"o{\ss}ler}
\author{Marcus Schmidt}
\affiliation{Max Planck Institute for Chemical Physics of Solids, 01067 Dresden, Germany}

\author{Oleg~Janson}
\affiliation{Institute for Theoretical Solid State Physics, Leibniz IFW Dresden, 01069 Dresden, Germany}

\author{Achyut Tiwari}
\affiliation{1. Physikalisches Institut, Universit{\"a}t Stuttgart, 70569 Stuttgart, Germany}

\author{Martin Dressel}
\affiliation{1. Physikalisches Institut, Universit{\"a}t Stuttgart, 70569 Stuttgart, Germany}

\author{Alexander A. Tsirlin}
\email{altsirlin@gmail.com}
\affiliation{Felix Bloch Institute for Solid-State Physics, University of Leipzig, 04103 Leipzig, Germany}


\begin{abstract}
Presence of magnetism in potentially altermagnetic RuO$_2$ has been a subject of intense debate. Using broadband infrared spectroscopy combined with density-functional band-structure calculations, we show that optical conductivity of RuO$_2$, the bulk probe of its electronic structure, is well described by the nonmagnetic model of this material. The sharp Pauli edge demonstrates the presence of a Dirac nodal line lying 45\,meV below the Fermi level. Good match between the experimental and \textit{ab initio} plasma frequencies underpins weakness of electronic correlations. The intraband part of the optical conductivity indicates Fermi-liquid behavior with two distinct scattering rates below 150\,K. Fermi-liquid theory also accounts for the temperature-dependent magnetic susceptibility of RuO$_2$ and allows a consistent description of this material as paramagnetic metal.
\end{abstract}

\maketitle


{\cdbl\textit{Introduction.}} 
Ruthenium dioxide is a simple inorganic solid that finds applications in catalysis~\cite{over2012}, microelectronics~\cite{iles1967}, and supercapacitors~\cite{majumdar2019}. The recent comeback of interest in this material has been triggered by the proposed realization of altermagnetism, thanks to the non-symmorphic nature of the RuO$_2$ crystal structure~\mbox{\cite{smejkal2022,feng2022,bose2022,bai2022,karube2022,jeong2024}}. Resonant x-ray scattering~\cite{zhu2019,gregory2022} and neutron diffraction experiments~\cite{berlijn2017} pinpointed the presence of ordered magnetic moments on the Ru atoms. Moreover, angle-resolved photoemission spectroscopy (ARPES) performed using circularly polarized soft x-rays demonstrated the splitting of spin-polarized bands expected in altermagnets~\cite{fedchenko2024}. On the other hand, magnetotransport measurements that allow a bulk probe of the Fermi surface~\cite{graebner1976} are consistent with the nonmagnetic band structure~\cite{mattheiss1976,yavorsky1996}. Moreover, recent muon spin relaxation~\cite{hiraishi2024} and renewed neutron diffraction experiments~\cite{kessler2024} do not find indications of bulk magnetism in RuO$_2$, thus leaving magnetic instability and, consequently, realization of altermagnetism in this material highly controversial.

Here, we report the optical spectroscopy study of RuO$_2$ that probes bulk electronic structure of this material across a broad range of energies, both above and below the Fermi level. Surprisingly, previous optical measurements mainly focused on the $p-d$ transitions above 2\,eV~\cite{goel1981,park1987,belkind1992,hones1995,mondio1997}, thus leaving the most interesting energy range covering Ru valence bands and their itinerant carriers unexplored. Using broadband infrared spectroscopy combined with density-functional calculations, we demonstrate that the optical response of RuO$_2$ is well described by the nonmagnetic band structure that perfectly reproduces several low-energy peaks arising from the optical transitions within the $t_{2g}$ bands, as well as the Pauli edge~\cite{pronin2021} indicative of the Dirac nodal lines located 45\,meV below the Fermi level. By comparing the experimental and calculated plasma frequencies, we show that RuO$_2$ should be classified as weakly correlated paramagnetic metal. The intraband part of its optical response manifests Fermi-liquid behavior, which is also consistent with the weak and continuous increase in the magnetic susceptibility with temperature. 
\smallskip

\begin{figure*}
  \centering
  \includegraphics[width=1\textwidth]{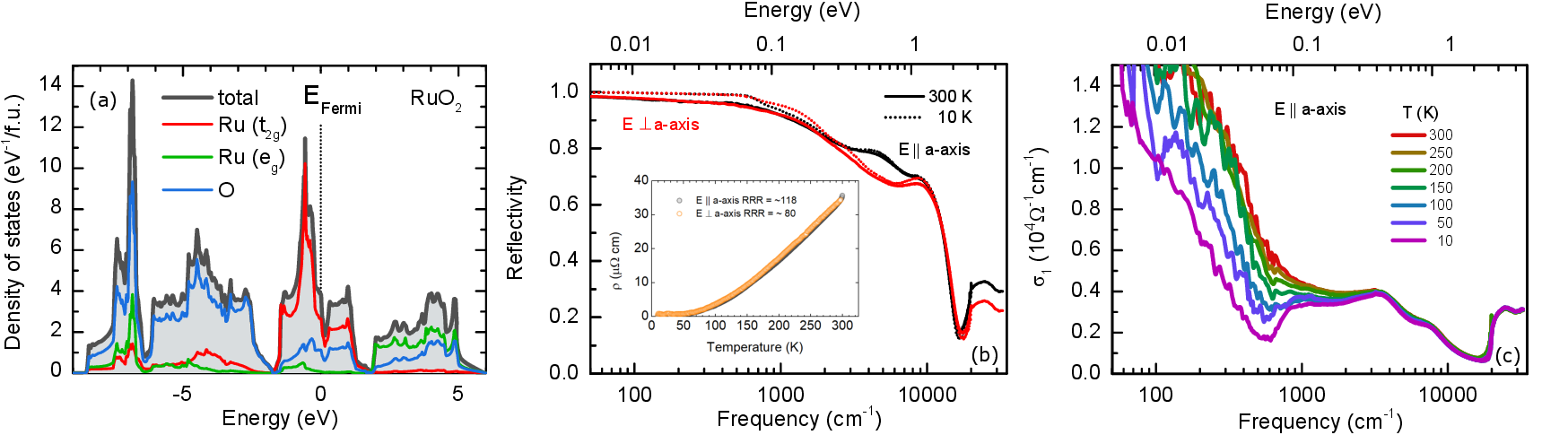}
    \caption{(a) Density of states for the nonmagnetic RuO$_2$. The states near the Fermi energy are dominated by the Ru $t_{2g}$ orbitals. (b) Broadband reflectivity measured along and perpendicular to the $a$-axis at room temperature and 10~K. No temperature dependence is observed, except at low energies where the intraband transitions are dominant. (c) Temperature-dependent in-plane optical conductivity. The inset in (b) shows the dc-resistivity of our RuO$_2$ single crystal.  }
    \label{F1}
\end{figure*}

{\cdbl\textit{Methods.}} Single crystals of RuO$_2$ have been grown by chemical vapor transport~\cite{supplement} and characterized by magnetic susceptibility as well as dc-electrical resistivity measurements [inset of Fig.~\ref{F1}(b)]. Apart from the Curie tail at low temperatures, magnetic susceptibility weakly increases on heating up to at least 400\,K, consistent with the previous reports~\cite{guthrie1931,fletcher1968,ryden1970b,berlijn2017}, whereas electrical transport shows the typical metallic behavior with the residual resistivity ratio (RRR) of 118, which is on par with the best samples used in transport measurements of bulk crystals~\cite{ryden1970,huang1982,lin2004,pawula2024} and almost two orders of magnitude higher than the typical RRR values of the thin-film samples~\cite{lin1999,uchida2020,ruf2021,feng2022,karube2022,liu2023}. 

Reflectivity measurements were performed with the Bruker Vertex80v and IFS113v spectrometers in a broad frequency range ($50-20000$ cm$^{-1}$) on as-grown single crystals [Fig.~\ref{F1}(b)]. The (011) surface was used for polarization-dependent measurements, thus allowing two nonequivalent directions, $E\,\|\,a$-axis and $E\!\perp\! a$-axis (mostly $E\,\|\,c$~\cite{supplement}). Optical conductivity was obtained by the Kramers-Kronig transformation of the reflectivity data. We used Hagen-Rubens extrapolations in the low-energy regime, whereas the high-energy region was extrapolated by using the ellipsometry data collected at room temperature up to 6 eV (unpolarized) and x-ray scattering functions above 6\,eV~\cite{Tanner2015}. The temperature-dependent spectra for the in-plane conductivity are shown in Fig.~\ref{F1}(c). The data for $E\perp a$ can be found in the Supplemental Material~\cite{supplement}.

Density-functional band-structure calculations were performed on the full-relativistic level in the Wien2K~\cite{wien2k,blaha2020} and FPLO~\cite{fplo} codes using the experimental crystal structure from Ref.~\cite{haines1997} and Perdew-Burke-Ernzerhof (PBE) exchange-correlation potential~\cite{pbe96}. The nonmagnetic band structure was obtained from PBE+SO calculations. Additionally, the mean-field DFT+$U$ correction was introduced for stabilizing the magnetic solution similar to the previous studies~\cite{ahn2019,smejkal2023,smolyanyuk2024,zhou2024,hariki2024}. Optical conductivity was obtained by the internal routine of Wien2K~\cite{draxl2006}.
\smallskip

\begin{figure}
  \centering
  \includegraphics[width=0.45\textwidth]{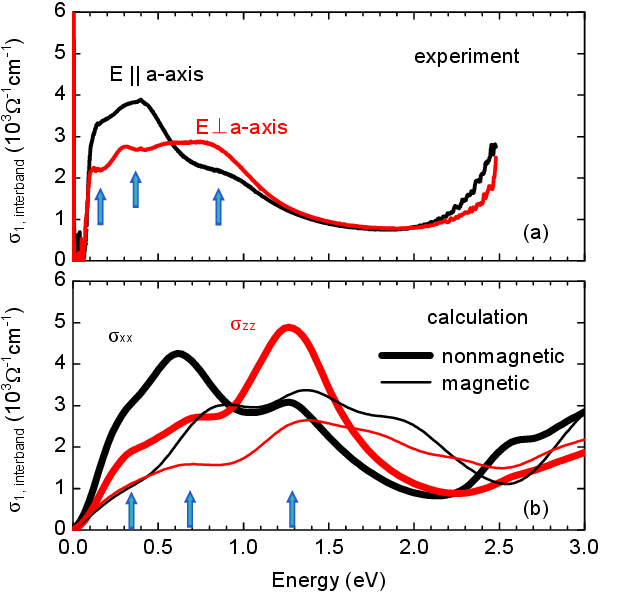}
    \caption{ (a) Experimental interband optical conductivity of RuO$_2$ for $E\,\|\,a$ and $E\!\perp\!a$ obtained by subtracting the intraband contribution from the spectra. The main absorption features are shown with arrows. (b) Calculated interband optical conductivity for the nonmagnetic band structure of RuO$_2$ (PBE+SO) and for the magnetic solution ($U$ = 2~eV, $J=0.3$\,eV). The 0.1~eV broadening is applied to the calculated spectra. Note that the experimental data for $E\!\perp\!a$ are dominated by the $E\,\|\,c$ signal~\cite{supplement}, thus showing the features characteristic of $\sigma_{zz}$, but those features are less pronounced than in the calculations. }
    \label{F2}
\end{figure}

{\cdbl\textit{Band structure and optical conductivity.}} 
Fig.~\ref{F1}(a) shows the density of states for nonmagnetic RuO$_2$. The states near the Fermi level are dominated by the Ru $t_{2g}$ orbitals and separated by small gaps from the Ru $e_g$ states above 2\,eV and from the O $2p$ states below $-2$\,eV. These separations are clearly visible in the optical response of the material. Indeed, calculated optical conductivity of the nonmagnetic RuO$_2$ [Fig.~\ref{F2}(b)] shows a pronounced minimum around 2\,eV that separates optical transitions within the $t_{2g}$ bands from the O $2p$ $\rightarrow$ Ru~$t_{2g}$ and Ru $t_{2g}$ $\rightarrow$ Ru $e_g$ transitions. 

Experimental interband optical conductivity obtained by subtracting the intraband response reveals a relatively weak anisotropy and clearly shows the minimum at $1.5-2.0$\,eV separating the transitions within the $t_{2g}$ manifold from the higher-energy transitions. Three main absorption features shown by arrows in Fig.~\ref{F2}(a) appear below this minimum. All of them can be traced in the calculated optical conductivity obtained from the nonmagnetic band structure. This nonmagnetic band structure correctly shows the crossing of $\sigma_{xx}$ and $\sigma_{zz}$ around 0.65\,eV (experiment) and 0.9\,eV (calculation), as well as the merge of $\sigma_{xx}$ and $\sigma_{zz}$ in the energy range of the minimum, around 2.0\,eV. 

The magnetic solution is unstable in PBE+SO and requires the addition of electronic correlations that change the positions of the Ru bands and shift the spectral weight to higher energies. The minimum in the optical conductivity appears above 2.5\,eV only and becomes less pronounced [Fig.~\ref{F2}(b)]. Adding magnetism clearly deteriorates the description of the experimental optical conductivity, as all features shift to much higher energies compared to the experiment, and the changes between the different polarization directions cannot be satisfactorily reproduced.

The best match between the experimental spectrum and the optical conductivity obtained from the nonmagnetic solution is achieved when the calculated energies are divided by the factor of 1.3. Such re-scaling is not uncommon in weakly correlated electronic systems where band energies are slightly renormalized compared to the band picture~\cite{uykur2022}. The magnitude of electronic correlations can be gauged by comparing the experimental and calculated plasma frequencies or the spectral weights~\cite{shao2020,wenzel2022,wenzel2024}. Our nonmagnetic calculation returns $\omega_p^{xx}=3.29$\,eV and $\omega_p^{zz}=3.45$\,eV in a very good agreement with the experimental values of 3.16 and 3.34\,eV for $E\,\|\,a$ and $E\!\perp\!a$, respectively. The ratio of the spectral weights of the intraband contribution, SW$_{\rm exp}$/SW$_{\rm DFT}$=0.87$\pm0.05$ (for both directions, but the comparison is more accurate for the in-plane direction), further identifies RuO$_2$ as a weakly correlated metal. 

The magnetic solution returns $\omega_p^{xx}=2.65$\,eV and $\omega_p^{zz}=2.58$\,eV and clearly underestimates plasma frequencies of RuO$_2$. We thus conclude that bulk RuO$_2$ can only be described on the level of standard band theory without including magnetism or the on-site Coulomb repulsion $U$. Our data suggest that the bulk electronic structure of RuO$_2$ is not consistent with that of an altermagnet. 
\smallskip

\begin{figure}
  \centering
  \includegraphics[width=0.4\textwidth]{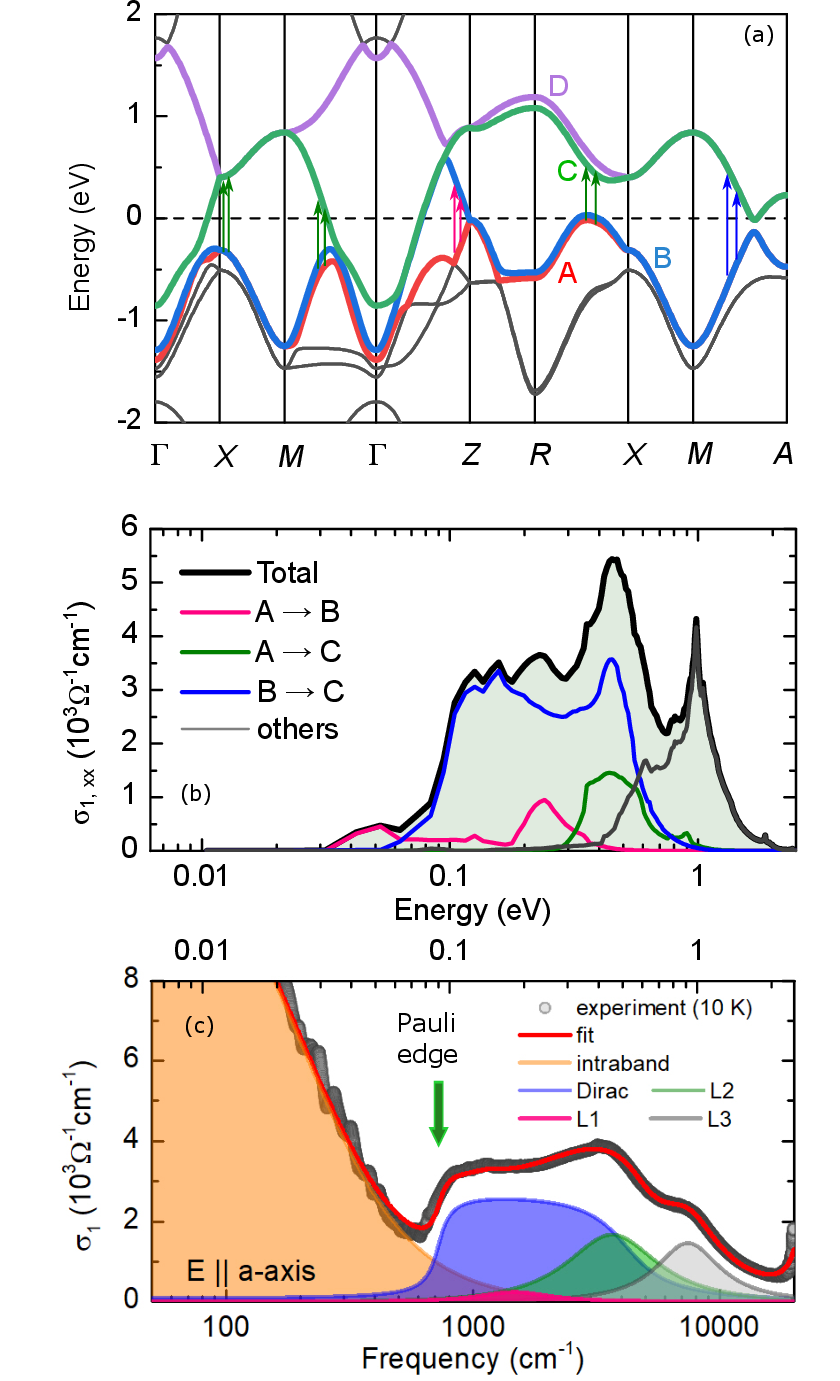}
    \caption{(a) Band structure of nonmagnetic RuO$_2$. Arrows indicate the major absorption features, which are color coded for the band-resolved optical conductivity. (b) Band-resolved optical conductivity with the energy axis divided by the factor of 1.3. No broadening is included into the calculation. Three major absorption features are shown with colored lines, whereas the absorption around 1~eV shown with the gray line is the combination of several interband transitions. (c) Experimental optical conductivity at 10~K and the decomposition of the spectrum into the intraband contribution and the interband contribution that consists of the Pauli edge and several Lorentzian peaks showing a remarkable agreement with the band-resolved optical conductivity from DFT. }
    \label{F3}
\end{figure}

{\cdbl\textit{Ru $t_{2g}$ bands.}} Previous optical studies of RuO$_2$ did not probe the energy range below $0.5-1.0$\,eV and could not fully resolve the optical transitions within the $t_{2g}$ bands~\cite{goel1981,krasovska1995,almeida2006}. With the data extending to low energies, we identify three main features in this region (Fig.~\ref{F3}). The sharp increase in the optical conductivity around 0.1\,eV followed by the almost flat region up to 0.4\,eV is reminiscent of the Pauli edge in Dirac semi-metals~\cite{pronin2021,schilling2017}. At higher energies, $\sigma_{xx}$ peaks at $0.5-0.6$\,eV and develops a visible shoulder around 1.0\,eV. 

Our band-resolved calculations show that the spectral weight below 0.4\,eV mostly originates from the bands $B$ and $C$ that form a Dirac nodal line, which ends at the gapped Dirac point along the $A-M$ direction. The sharp Pauli edge observed in the interband optical conductivity around 0.1\,eV is a direct experimental fingerprint of this topological feature~\cite{ahn2019,jovic2018} that should lead to the large spin-Hall effect in RuO$_2$~\cite{sun2017}. Other possible low-energy transitions, for example between the linear bands $A$ and $B$ along $Z-\Gamma$, give only a small contribution in this energy range. Above 0.4\,eV, optical transitions along the $X-R$ line start contributing to the spectrum, resulting in the peak of the optical conductivity. The shoulder around 1.0\,eV is due to multiple interband transitions and reflects the fact that several bands develop almost flat regions at $-0.5$\,eV and $+0.5$\,eV, respectively. 

The interband part of the experimental spectrum is decomposed using the Pauli edge and three Lorentzian features [Fig.~\ref{F3}(c)] that show a striking agreement with the individual contributions from the band-resolved analysis. Our experimental data put the Pauli edge to 90~meV and pin the Dirac nodal line of bulk RuO$_2$ to 45~meV below the Fermi level.  
\smallskip

\begin{figure}
  \centering
  \includegraphics[width=0.5\textwidth]{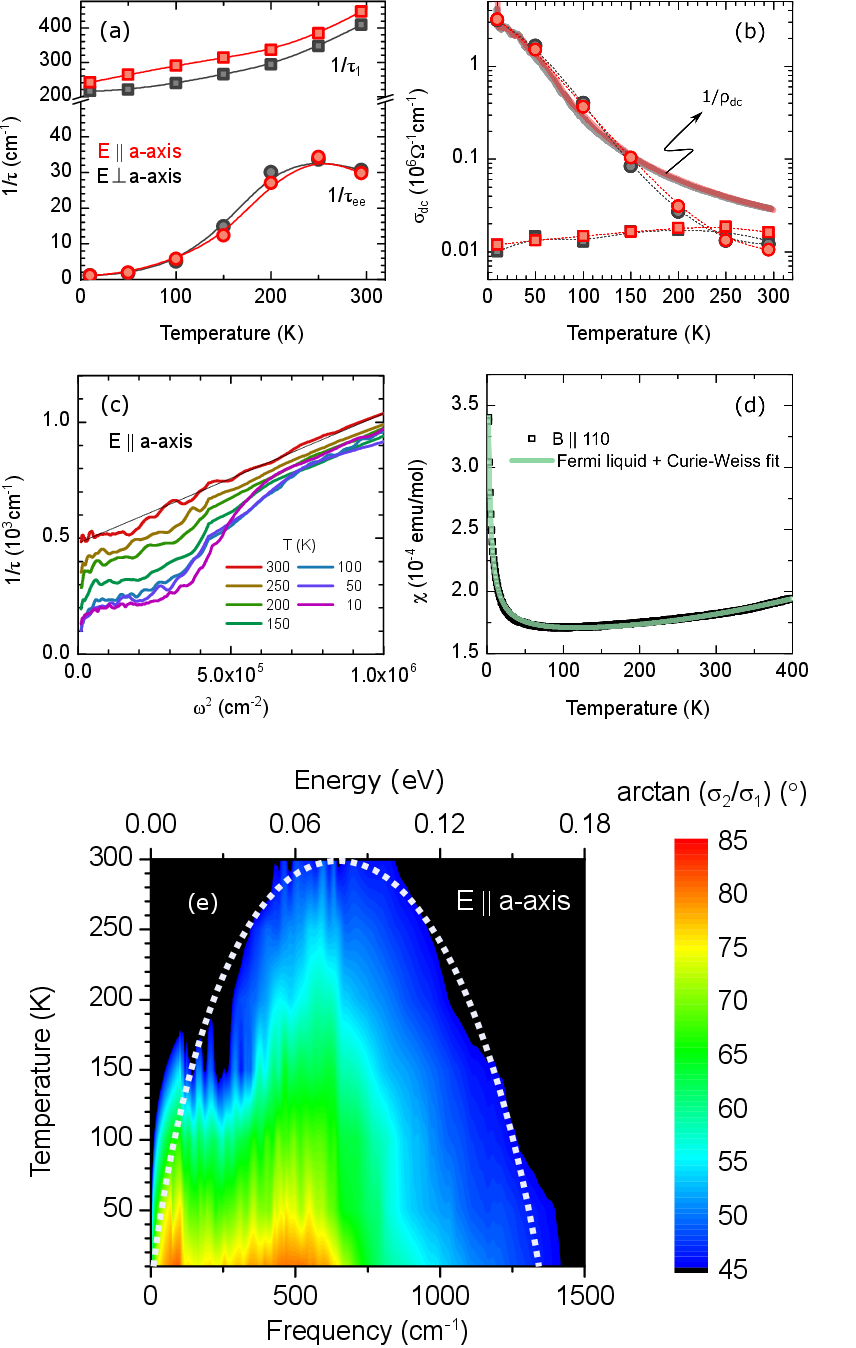}
  \caption{ (a) Temperature-dependent scattering rates of the intraband contribution and the corresponding dc-conductivities overlayed with the values obtained from the dc-resistivity values (b). (c) Quadratic frequency dependence of the scattering rate of the extended Drude model. (d) Temperature-dependent magnetic susceptibility of RuO$_2$ measured in the applied field of 1\,T and its fit using the $T^2$ term of the paramagnetic metal along with the Curie-like impurity contribution~\cite{supplement}. (e) The phase angle $\sigma_2/\sigma_1$ determined from our optical experiments in the $T - \omega$ domain. The semi-elliptical region of $\sigma_2>\sigma_1$ witnesses the Fermi-liquid behavior. }
    \label{F4}
\end{figure}

{\cdbl\textit{Intraband contribution.}} 
Having elucidated the interband part of the optical conductivity, we now discuss the optical response of itinerant carriers. Interestingly, this response can not be modeled with the conventional Drude peak. Instead, our data reveal the presence of two distinct scattering rates that both increase on heating [Fig.~\ref{F4}(a)]. The respective contributions to the dc-conductivity shown in Fig.~\ref{F4}(b) suggest that at low temperatures the dc-conductivity of RuO$_2$ is mainly determined by the lower relaxation rate, whereas above 200\,K the contributions of both relaxation rates are comparable. This situation is reminiscent of an interplay between phonon scattering and electron-electron scattering, with the former becoming predominant at elevated temperatures. A similar scenario was also described theoretically for a system with two bands and the interband electron-electron scattering~\cite{maslov2016}.

Alternatively, the intraband part of the optical conductivity can be analyzed using the extended Drude model with the frequency-dependent relaxation rate, $1/\tau(\omega)$~\cite{basov2011,berthod2013}. This analysis reveals the quadratic frequency dependence typical of a Fermi liquid [Fig.~\ref{F4}(c)]. Below 150\,K, the difference between the two scattering rates becomes prominent, and a step in the frequency dependence appears. The quadratic behavior is nevertheless preserved at frequencies both above and below the step. The Fermi-liquid behavior is further witnessed by the phase angle of the imaginary and real parts of the optical conductivity, $\sigma_2/\sigma_1$, plotted in the $T-\omega$ domain in Fig.~\ref{F4}(e). The semi-elliptical shape of the region where $\sigma_2>\sigma_1$ is typical of the Fermi-liquid regime~\cite{tytarenko2015,pustogow2021}.

\smallskip


{\cdbl\textit{Discussion.}} 
Our results suggest that the bulk electronic structure of RuO$_2$ can be well understood on the level of band theory without including correlations or magnetism. The analysis of the intraband spectral weight gives direct evidence for the weakness of electronic correlations in this material. It further supports the nonmagnetic nature of bulk RuO$_2$ because the addition of the on-site Coulomb repulsion is essential for stabilizing magnetic moments on the Ru atoms. Adding magnetism deteriorates the agreement with the experiment for both intraband and interband parts of the spectrum.

Possible altermagnetism of RuO$_2$ has been ascribed to a Fermi surface instability that mostly arises from the small pockets of the Fermi surface along $X-R$ and \mbox{$A-M$}~\cite{berlijn2017,ahn2019}. Minor changes in the position of the Fermi level, e.g., by a variation of the oxygen content, should significantly affect these parts of the Fermi surface and may thus change the proclivity of RuO$_2$ for magnetism. This renders the oxygen nonstoichiometry scenario~\cite{smolyanyuk2024} a viable explanation for the presence of magnetism in thin films and absence of magnetism in the bulk, especially in the light of the drastically low RRR values reported for the thin-film samples~\cite{lin1999,uchida2020,ruf2021,feng2022,karube2022,liu2023}. Optical spectroscopy would be a useful tool for probing the true electronic structure of such thin films. In particular, the position of the Pauli edge can serve as a sensitive probe of the Fermi level.

In the absence of magnetism, bulk RuO$_2$ should be thought of as paramagnetic metal. Indeed, intraband part of its optical conductivity reveals the Fermi-liquid type of behavior with the quadratic scaling of the scattering rate. This finding justifies the application of Fermi-liquid theory also to the magnetic properties of bulk RuO$_2$. Temperature-dependent susceptibilities of paramagnetic transition metals have been previously described using the $T^2$ term that arises in Fermi-liquid theory~\cite{misawa1976,carneiro1977,maslov2003} as well as in band theory~\cite{kriessman1954,shimizu1981}. Fig.~\ref{F4}(d) shows that a similar description is possible in the case of RuO$_2$ too when the low-temperature upturn is taken into account by adding the Curie-like contribution that corresponds to only 0.17\% of \mbox{spin-$\frac12$} impurities~\cite{supplement}. Therefore, the monotonic increase in the magnetic susceptibility of RuO$_2$ is not an indication of antiferromagnetism. It manifests the paramagnetic response of itinerant carriers, similar to simple $4d$ transition metals such as Ru and Rh~\cite{galoshina1974}.

Our optical data not only support the absence of magnetism in bulk RuO$_2$, as inferred from various probes of magnetic order~\cite{hiraishi2024,kessler2024}, but also give strong evidence for the weakness of electronic correlations in this material. These data set an important constraint on stabilizing magnetism within the RuO$_2$ framework. The apparent signatures of altermagnetism reported in thin-film samples~\cite{feng2022,bose2022,bai2022,karube2022,jeong2024,fedchenko2024} suggest that effects like strain and off-stoichiometry may render RuO$_2$ magnetic. However, this putative magnetism has to be assessed without invoking sizable electronic correlations, contrary to the existing theoretical studies that typically enforce magnetism within a correlated scenario.

\smallskip


{\cdbl\textit{Conclusions.}} 
Optical properties of RuO$_2$ are well described by the nonmagnetic band structure and suggest the absence of altermagnetism in bulk ruthenium dioxide. Interband part of the optical conductivity reveals several absorption peaks due to optical transitions within the Ru $t_{2g}$ bands along with the Pauli edge that manifests the Dirac nodal line located 45\,meV below the Fermi level. The analysis of the plasma frequencies and intraband spectral weight classifies RuO$_2$ as a weakly correlated metal. Its intraband optical response abides by the Fermi-liquid theory, which also explains the monotonic increase in the magnetic susceptibility and offers a consistent description of bulk RuO$_2$ as paramagnetic metal.

\acknowledgments
We are grateful to Vicky Hasse for technical assistance with crystal growth and Gabriele Untereiner for preparing single crystals for optical measurements. M.W. is supported by Center for Integrated Quantum Science and Technology (IQST) Stuttgart/Ulm via a project funded by the Carl Zeiss Stiftung. The work in HZDR is funded by the DFG UY 63/5-1. The work in Leipzig was funded by the Deutsche Forschungsgemeinschaft (DFG, German Research Foundation) -- TRR 360 -- 492547816 (subproject B1). 

\bibliography{RuO2-optics}

\newpage
\makeatletter
\renewcommand\@bibitem[1]{\item\if@filesw \immediate\write\@auxout
    {\string\bibcite{#1}{S\the\value{\@listctr}}}\fi\ignorespaces}
\def\@biblabel#1{[S#1]}
\makeatother
\setcounter{figure}{0}
\renewcommand{\thefigure}{S\arabic{figure}}
\newcommand{\av}{\mathbf a}
\newcommand{\bv}{\mathbf b}
\newcommand{\cv}{\mathbf c}

\clearpage
\newpage
\onecolumngrid
\begin{center}
\large{\textbf{\textit{Supplemental Material} \\ Fermi-liquid behavior of non-altermagnetic RuO$_2$}}\\
\end{center}
\begin{center}
\large{Maxim Wenzel$^1$, Ece Uykur$^{2,*}$, Sahana Rößler$^3$, Marcus Schmidt$^3$,\\
Oleg Janson$^4$, Achyut Tiwari$^1$, Martin Dressel$^1$, and Alexander A. Tsirlin$^{5,\dagger}$}\\
\end{center}
\begin{center}
\begin{small}
$^1$\textit{1. Physikalisches Institut, Universität Stuttgart, 70569 Stuttgart, Germany}\\
$^2$\textit{Helmholtz-Zentrum Dresden-Rossendorf, Inst Ion Beam Phys \& Mat Res, D-01328 Dresden, Germany}\\
$^3$\textit{Max Planck Institute for Chemical Physics of Solids, 01067 Dresden, Germany}\\
$^4$\textit{Institute for Theoretical Solid State Physics, Leibniz IFW Dresden, 01069 Dresden, Germany}\\
$^5$\textit{Felix Bloch Institute for Solid-State Physics, University of Leipzig, 04103 Leipzig, Germany}\\
\end{small}
\end{center}
\begin{center}

\end{center}

\twocolumngrid
\subsection{Sample preparation and characterization}
Single crystals of RuO$_2$ were grown by a chemical transport reaction following Ref.~\hyperlink{cite}{\color{blue}[S1]}. Commercial microcrystalline RuO$_2$ powder (Aldrich 99.9\%) was used as the starting material. Chemical transport reaction was performed in the temperature gradient from 1090\,$^{\circ}$C (source) to 1000\,$^{\circ}$C (sink) with an addition of 3.5 mg/cm$^3$ TeCl$_4$ (Aldrich 99\%) and 5 mg/cm$^3$ Au$_2$O$_3$ (Alfa Aesar 99.99\%) as supplementary oxygen source. The growth was performed in an evacuated quartz glass ampoule with an additional quartz insert that contained the source material and prevented its reaction with the ampoule wall. The transport additives were placed in an open capillary so that RuO$_2$ does not come in contact with gold formed upon thermal decomposition of Au$_2$O$_3$. 

RuO$_2$ can be transported with the addition of TeCl$_4$ or, alternatively, with oxygen~\hyperlink{cite}{\color{blue}[S2]}. The combination of transport additives and especially the increase in the oxygen partial pressure via the added Au$_2$O$_3$ contributes to a higher transport rate and the deposition of stoichiometric RuO$_2$ crystals, which are free from oxygen deficiency. The growth was terminated after 10 days by quenching the ampoule in cold water. Some of the obtained crystals had regular shape with the maximum size of about $1\times 1\times 1$\,mm$^3$.

X-ray powder diffraction (Huber G670 camera, CuK$_{\alpha1}$ radiation) measured on several crushed single crystals confirmed the absence of any impurity phases. Le Bail fit of the XRD pattern returns the lattice parameters $a=4.4904(1)$\,\r A and $c=3.1062(1)$\,\r A in an excellent agreement with the values reported for the stoichiometric RuO$_2$ samples~\hyperlink{cite}{\color{blue}[S3, S4]}.

\subsection{Magnetic susceptibility} 

Magnetic susceptibility was measured with MPMS\,3 from Quantum Design in the applied field of 1\,T. Single crystal was glued to a quartz sample holder. No difference between the field-cooling and zero-field-cooling regimes was observed. Field-dependent magnetization curves recorded at several constant temperatures showed linear behavior expected for paramagnets. Below 10\,K, the magnetization curves showed bending caused by saturation of paramagnetic spins.

Temperature-dependent magnetic susceptibility of RuO$_2$ was fitted with the equation
\begin{equation}
\chi(T) =\chi_0 + A\,T^2 + \frac{C}{T-\Theta}
\label{MFL}
\end{equation}
where $\chi_0$ is the temperature-independent contribution, $A\,T^2$ describes temperature dependence of the susceptibility for a paramagnetic metal (Fermi liquid), whereas $C/(T-\theta)$ accounts for the low-temperature upturn caused by impurity spins. 

The fit returned $\chi_0 = 1.63 \times 10^{-4}$ emu/mol, $A = 1.86 \times 10^{-10}$\,emu\,mol$^{-1}$\,K$^{-2}$, $C = 6.3 \times 10^{-4}$\,emu\,K/mol, and $\Theta = -1.5$\,K. The $\Theta$ value close to zero confirms paramagnetic nature of the impurity spins, whereas their amount (0.17\%) can be quantified by a comparison to the Curie constant of $C=(g\mu_B)^2S(S+1)/3k_B=0.375$\,emu\,K/mol for $S=\frac12$ and $g=2$. 


\subsection{Details of the broadband infrared measurements}

\begin{figure*}
\centering
  \includegraphics[width=1\textwidth]{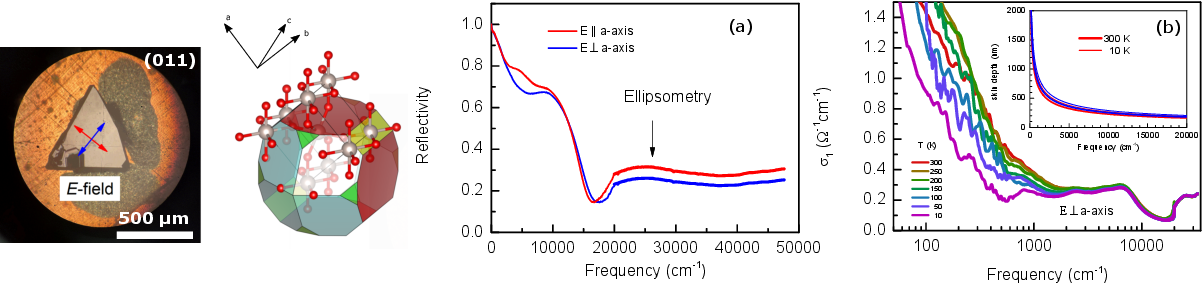}
    \caption{ RuO$_2$ single crystal used in polarization-dependent measurements. The crystallographic axes are shown with the arrows, and the representative facet is illustrated using VESTA~\textcolor{blue}{[S5]}. (a) Room-temperature reflectivity spectra for $E\parl a$ and $E\per a$ are color-coded according to the crystal picture. The bold lines at higher energies are the unpolarized ellipsometric measurement, which is normalized to our reflectivity data prior to the Kramers-Kronig transformation. (b) Temperature-dependent optical conductivity for $E\per a$ (the temperature-dependent $E\parl a$ data are shown in the main text).  }
    \label{S1}
\end{figure*}

Broadband reflectivity measurements were performed with the Bruker Vertex80v coupled with the Hyperion infrared microscope in the frequency range of $600-20000$~\cm\ and with the IFS113v spectrometer at $50-700$~\cm. An as-grown sample (as shown in Fig.~\ref{S1}) used for the measurements exhibits a roughly triangular facet with the lateral size of $\sim$~500~$\mu$m and the sample thickness of 200-300~$\mu$m. Polarization-dependent measurements were performed on this (011) surface with two polarization directions shown in Fig.~\ref{S1}. One of the direction is the crystallographic $a$-axes ($E\parl a$), whereas the second direction, $E\per a$, is $[0\bar 11]$ with a significant $c$-component.

Optical conductivity was obtained by the Kramers-Kronig transformation of the reflectivity data. Due to the highly metallic nature of the samples, we used Hagen-Rubens extrapolations in the low-energy regime. For the high-energy part there seemed to be a strong absorption feature right above the measurement limit of our FTIR spectrometers. Therefore, we employed an ellipsometric measurement (Fig.~\ref{S1}(a)) at room temperature up to $\sim$~6~eV in an unpolarized configuration as a bridge function to the x-ray scattering functions employed at even higher frequencies. Temperature dependence of the optical conductivity for $E\per a$ is shown in Fig.~\ref{S1}(b), whereas the $E\parl a$ data are given in the main text. The overall behavior of the spectra is similar for both polarization directions. A slightly larger intraband contribution is seen for $E\per a$, as also reflected by the higher value of the plasma frequency, in agreement with our DFT calculations (see the main text). 

We also estimated the skin depth of the sample in the broad frequency range as given in Fig.~\ref{S1}(b) (inset). The skin depth changes from 200 nm at higher energies to $\sim$~ 2.5~$\mu$m at around 10 meV suggesting that bulk electronic structure of RuO$_2$ is probed in our experiments. There are no significant changes between the different polarization directions and temperatures (the skin depth slightly increases with decreasing temperature). \\


\subsection{Plasma frequency and gauge of electronic correlations}

\begin{figure}
\centering
  \includegraphics[width=0.4\textwidth]{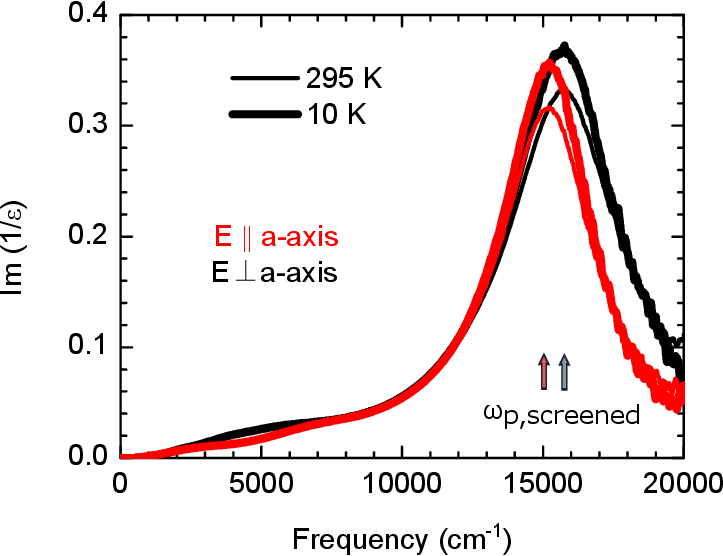}
    \caption{Loss function vs. frequency at room temperature (thin lines) and at 10~K (thick lines) for both polarization directions. Screened plasma frequencies are shown with the arrows. The red and black colors correspond to the $E\parl a$ and $E\per a$, respectively. The out-of-plane contribution leads to a slightly higher plasma frequency for $E\per a$. No appreciable temperature dependence is observed for either of the polarization directions. }
    \label{S2}
\end{figure}

Plasma frequency can be estimated by different methods. The zero-crossing of the real part of the dielectric function reflects the screened plasma frequency, which should be renormalized by the $\epsilon_\infty$, the high-energy permittivity, using $\omega_{p} =\omega_{p,{\rm screened}}\times \sqrt{\epsilon_\infty}$.  The screened plasma frequency can also be estimated from the loss function, which is shown in Fig.~\ref{S2}. The slightly higher plasma frequency is observed for E$\per{}a$ compared to E$\parl a$. On the other hand, neither crystallographic direction shows any temperature dependence, indicating that the overall plasma frequency ($n_e/m^*$) is conserved. The $\epsilon_{\infty}$ values determined from the real part of the dielectric function are 2.83 and 2.97\,eV for E$\parl a$ and  E$\per a$, respectively. The plasma frequencies are then estimated as 3.16 and 3.34\,eV for E$\parl a$ and  E$\per a$, respectively.

Alternatively, plasma frequency can be obtained directly from the spectral weight (SW) of the intraband transitions. The area under $\sigma_1(\omega)$ is evaluated with a cutoff frequency [$\int_0^{\omega_{\rm cutoff}}\sigma_1(\omega)d\omega$] that marks the upper bound of the intraband transitions. This method becomes less accurate at higher temperatures because the intraband and interband transitions overlap. Therefore, we determined the SW by integrating the intraband contribution obtained by a decomposition of the experimental optical conductivity within the Drude-Lorentz approach. This way, we could also separate the contributions with the different scattering rates as explained in the main text. Eventually, independent of the chosen method, the obtained plasma frequencies for E$\parl a$ and E$\per a$ are consistent and further show a good agreement with the nonmagnetic DFT calculations. 

Correlation strength was gauged using the phenomenological approach of Ref.~\hyperlink{cite}{\color{blue}[S6]} that identifies the material as correlated if the ratio of its experimental and calculated plasma frequencies or spectral weights is well below 1.0. Our data return the ratio of 0.87$\pm$0.05 for both directions. The results for $E\parl a$ are more accurate because this geometry directly corresponds to the calculated $\sigma_{xx}$.


\subsection{Extended Drude analysis} 
In the classical treatment of the Drude-Lorentz formalism, the scattering rate of the Drude intraband contribution is defined as a constant, single frequency (1/$\tau$). On the other hand, in real materials, various inelastic channels can contribute to the scattering and in principle a frequency-dependent scattering rate is often observed. 

\begin{figure}
\centering
  \includegraphics[width=0.5\textwidth]{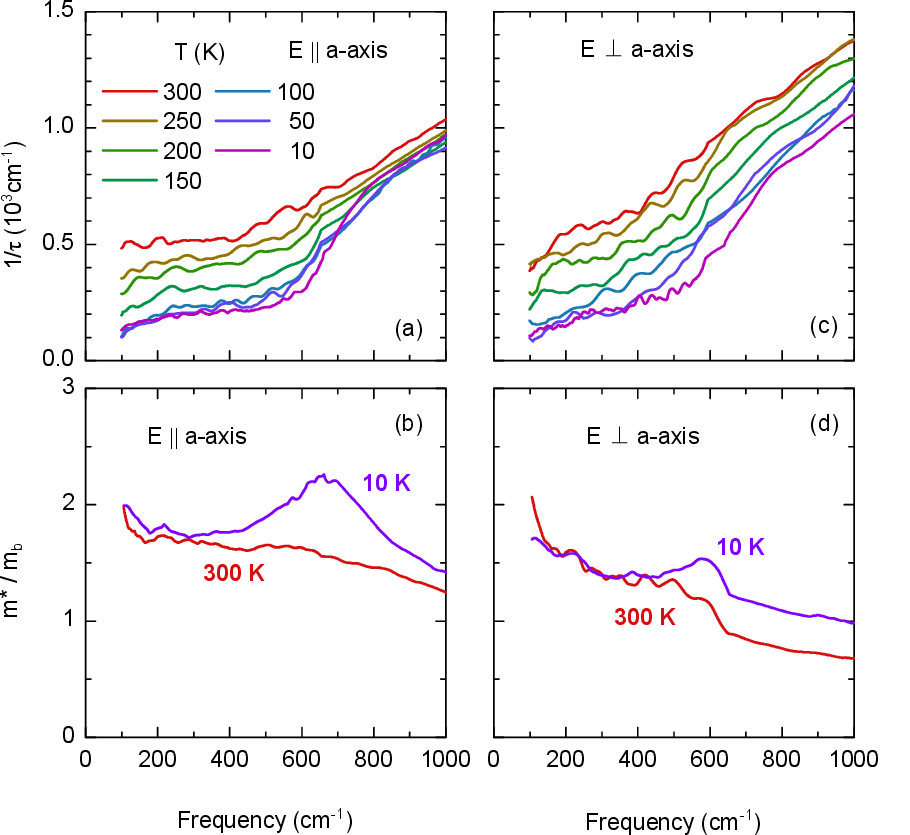}
    \caption{Results of the extended Drude analysis. (a,b) Frequency-dependent scattering rate and effective mass renormalization for $E\parl a$. (c,d) The same for $E\per a$.   }
    \label{S3}
\end{figure}

Extended Drude formalism takes these frequency-dependent effects into account by redefining the effective mass and scattering rate as follows,
\begin{equation}
\frac{m^*(\omega)}{m_b}=-\frac{\omega_p^2}{4 \pi \omega}\times {\rm Im}\left[\frac{1}{\tilde{\sigma}(\omega)}\right]
\end{equation}
\begin{equation}
\frac{1}{\tau(\omega)}=\frac{\omega_p^2}{4 \pi}\times {\rm Re}\left[\frac{1}{\tilde{\sigma}(\omega)}\right]
\end{equation}

Here, $\tilde{\sigma}(\omega)$ is the complex optical conductivity, $\omega_p$ is the plasma frequency, and $m_b$ is the band mass. The frequency-dependent scattering rates and mass enhancements are shown in Fig.~\ref{S3} for both $E\parl a$ and $E\per a$. The mass enhancement is comparable to the general noble metals~\hyperlink{cite}{\color{blue}[S7]}, whereas is quite small compared to typical correlated systems~\hyperlink{cite}{\color{blue}[S8]}/heavy fermions~\hyperlink{cite}{\color{blue}[S9, S10]} and further indicates the weakness of electronic correlations in RuO$_2$.


\subsection{Intraband transitions}
\begin{figure}
\centering
  \includegraphics[width=0.5\textwidth]{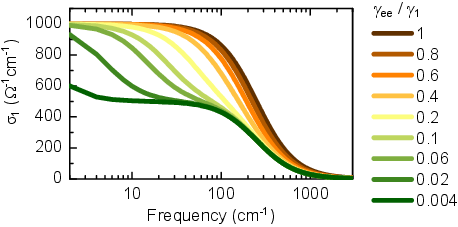}
    \caption{ Frequency dependent intraband optical conductivity. Scattering rate (where $\gamma = 1/\tau$) ratio is changed gradually. When the $\tau_{ee}$ is comparable to $\tau_1$ system can be simulated by a single Drude. Otherwise, a step like behavior is clearly observed. For the simulations $\sigma_{1,ee}$ and $\sigma_{1,1}$ are chosen to be equal and 500 $\Omega^{-1} cm^{-1}$ }
    \label{S4}
\end{figure}

Different scattering rates had to be taken into account in the analysis of the intraband part of the optical conductivity. These different scattering channels are witnessed by the step-like behavior of the optical conductivity. In Fig.~\ref{S4}, we simulate this behavior for two scattering channels with the different ratios of the scattering rates, $\gamma_{ee} / \gamma_1$ ($\gamma = 1/\tau$). In the simulation, the dc-conductivity values are kept the same, but the similar step-like-behavior and the related discussion also hold in the case of different dc-conductivity values.

We used a similar approach to fit the intraband contribution to the experimental optical conductivity. With the following representations, the low-energy part of $\sigma_1(\omega)$ and the step-like behavior can be well reproduced. With these fits, we also obtained the temperature dependence of $\tau_{ee}$ and $\tau_1$ and their relative contributions to the dc-conductivity as given in the main text (Fig. 4).

\begin{equation}
\gamma = \frac{1}{\tau} = \frac{1}{\tau_1} + \frac{1}{\tau_{ee}}
\end{equation}
\begin{equation}
\sigma_{1,{\rm dc}} = \sigma_{1,ee} + \sigma_{1,1}
\end{equation}
\begin{equation}
\sigma_{1,{\rm intraband}} = \frac{\gamma^2\, \sigma_{1,{\rm dc}}}{\omega^2+\gamma^2}
\end{equation}


\subsection{Additional DFT results}

Density-functional (DFT) calculations were performed with the $24\times 24\times 32$ $k$-mesh that comprised 2496 points in the irreducible part of the first Brillouin zone. This $k$-mesh allowed the convergence of the Fermi energy within 2\,meV and an excellent convergence of the optical conductivity at energies down to 0.01\,eV. 

\begin{figure}
\includegraphics{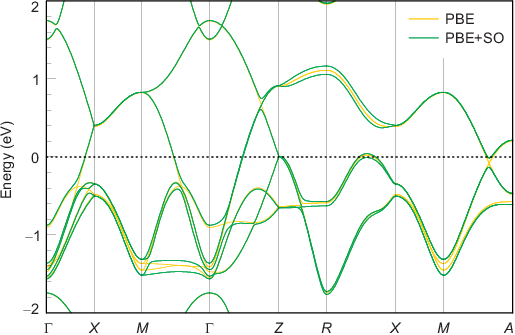}
\caption{\label{fig:bands}
Comparison of scalar-relativistic (PBE) and full-relativistic (PBE+SO) band structures of RuO$_2$. Both solutions are nonmagnetic.
}
\end{figure}

Fig.~\ref{fig:bands} shows the nonmagnetic band structure of RuO$_2$ calculated in the scalar-relativistic (PBE) and full-relativistic (PBE+SO) modes. Spin-orbit coupling causes only minor changes in the band dispersions. Most notably, it gaps out the Dirac point along $M-A$. Optical conductivity was obtained from full-relativistic calculations. Scalar-relativistic calculations lead to qualitatively similar results.

Spin-polarized PBE+SO calculations converge to a nonmagnetic solution. Therefore, correlation effects are required for stabilizing magnetism in bulk RuO$_2$. Such correlation effects are typically introduced within the mean-field DFT+$U$ procedure where values of the on-site Coulomb repulsion $U_d$ in excess of 1\,eV are required to create magnetic moments on the Ru atoms~\hyperlink{cite}{\color{blue}[S11, S12]}. The exact threshold for $U_d$ depends on the implementation of DFT+$U$, including the double-counting correction that can be calculated in the fully localized limit (FLL) or using the around-mean-field (AMF) approach.

\begin{figure}
\includegraphics{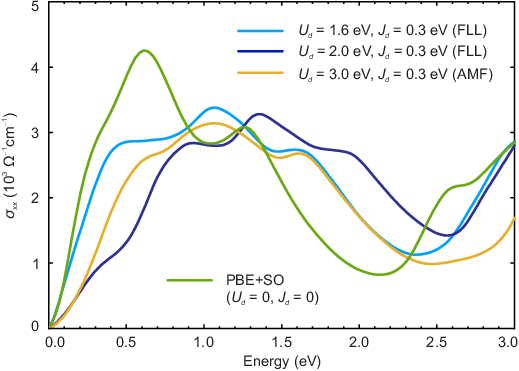}
\caption{\label{fig:sigma}
In-plane optical conductivity calculated in PBE+SO (nonmagnetic) and in PBE+SO+$U$ (magnetic) with different choices of the $U_d$ parameter and double-counting correction. Adding magnetism and correlations shifts all spectral features toward higher energies, at odds with the experimental data.}
\end{figure}

Optical conductivity was calculated in Wien2K. To this end, we chose $U_d=2.0$\,eV and $J_d=0.3$\,eV (FLL) as the typical DFT+$U$ parameterization used in the previous studies~\hyperlink{cite}{\color{blue}[S13, S14]}. This solution corresponds to the Ru magnetic moment of 0.83\,$\mu_B$ (note that this magnetic moment should not be directly compared with the FPLO value because of the different basis set). Fig.~\ref{fig:sigma} shows the in-plane optical conductivity for this and two other solutions that represent the lowest magnetic moments stabilized in FLL (0.60\,$\mu_B$, $U_d=1.6$\,eV, $J_d=0.3$\,eV) and AMF (0.53\,$\mu_B$, $U_d=3$\,eV, $J_d=0.3$\,eV). All three solutions show a significant upward shift of the spectrum compared to PBE+SO and, therefore, a quite poor agreement with the experimental optical conductivity. The shift of the spectrum is mainly controlled by the $U_d$ value that affects relative positions of the $t_{2g}$ bands. Our experimental data exclude such a shift in RuO$_2$ and, therefore, strongly support the uncorrelated and nonmagnetic band structure obtained on the PBE+SO level. 

\section*{Supplementary references}
\begin{small}

\begin{enumerate}[label={$\left[S\arabic*\right]$}]
\hypertarget{cite}  \item H. Oppermann and M. Ritschel, Zum chemischen Transport der \"{U}bergangsmetalldioxide mit Tellurhalogeniden, \href{https://doi.org/10.1002/crat.19750100504} {Krist. Tech.  \textbf{10}, 485 (1975)}.
\item M. Binnewies, R. Glaum, M. Schmidt, and P. Schmidt,  \href{https://doi.org/10.1515/9783110254655} {Chemical Vapor Transport Reactions (2012)}.
\item  Y. S. Huang, H. L. Park, and F. H. Pallak, Growth and characterization of RuO$_2$ single crystals, \href{https://doi.org/10.1016/0025-5408(82)90166-0} {Mater. Res. Bull. \textbf{17}, 1305 (1982)}.
\item J. Haines, J. M. L\'{e}ger, O. Schulte, and S. Hull, Neutron diffraction study of the ambient-pressure, rutile-type and the high-pressure, CaCl$_2$-type phases of ruthenium dioxide,\href{https://doi.org/10.1107/S0108768197008094} {Acta Cryst. \textbf{B53}, 880 (1997)}.
\item K. Momma and F. Izumi, VESTA 3 for three dimensional visualization of crystal, volumetric and morphology data, \href{https://doi.org/10.1107/S0021889811038970} {J. Appl. Crystallogr. \textbf{44}, 1272 (2011)}.
\item Y. Shao, A. N. Rudenko, J. Hu, Z. Sun, Y. Zhu, S. Moon, A. J. Millis, S. Yuan, A. I. Lichtenstein, D. Smirnov, Z. Q. Mao, M. I. Katsnelson, and D. N. Basov, Electronic correlations in nodal-line semimetals, \href{https://doi.org/10.1038/s41567-020-0859-z} {Nature Phys. \textbf{16}, 636 (2020)}.
\item S. J. Youn, T. H. Rho, B. I. Min, and K. S. Kim, Extended Drude model analysis of noble metals, \href{ https://doi.org/10.1002/pssb.200642097} {physica status solidi (b) \textbf{244}, 1354 (2007)}.
\item D. N. Basov, R. D. Averitt, D. van der Marel, M. Dressel, and K. Haule, Electrodynamics of correlated electron materials, \href{https://doi.org/10.1103/RevModPhys.83.471} {Rev. Mod. Phys. \textbf{83}, 471 (2011)}.
\item A. M. Awasthi, L. Degiorgi, G. Gr\"{u}ner, Y. Dalichaouch, and M. B. Maple, Complete optical spectrum of CeAl$_3$, \href{https://doi.org/10.1103/PhysRevB.48.10692} {Phys. Rev. B \textbf{48}, 10692 (1993)}.
\item G. Boss\'{e}, L. S. Bilbro, R. V. Aguilar, L. Pan, W. Liu, A. V. Stier, Y. Li, L. H. Greene, J. Eckstein, and N. P. Armitage, Low energy electrodynamics of the kondo-lattice antiferromagnet CeCu$_2$Ge$_2$, \href{https://doi.org/10.1103/PhysRevB.85.155105} { Phys. Rev. B \textbf{85}, 155105 (2012)}.
\item K.-H. Ahn, A. Hariki, K.-W. Lee, and J. Kune\v{s}, Antiferromagnetism in RuO$2$ as d-wave Pomeranchuk instability, \href{https://doi.org/10.1103/PhysRevB.99.184432} {Phys. Rev. B \textbf{99}, 184432 (2019)}.
\item A. Smolyanyuk, I. I. Mazin, L. Garcia-Gassull, and R. Valent\'{i}, Fragility of the magnetic order in the prototypical altermagnet RuO$_2$, \href{https://doi.org/10.1103/PhysRevB.109.134424} {Phys. Rev. B \textbf{109}, 134424 (2024)}.
\item A. Bose, N. J. Schreiber, R. Jain, D.-F. Shao, H. P. Nair, J. Sun, X. S. Zhang, D. A. Muller, E. Y. Tsymbal, D. G. Schlom, and D. C. Ralph, Tilted spin current generated by the collinear antiferromagnet ruthenium dioxide, \href{https://doi.org/10.1038/s41928-022-00744-8} {Nature Electronics \textbf{5}, 267 (2022)}.
\item X. Zhou, W. Feng, R.-W. Zhang, L. \v{S}mejkal, J. Sinova, Y. Mokrousov, and Y. Yao, Crystal thermal transport in altermagnetic RuO$_2$, \href{https://doi.org/10.1103/PhysRevLett.132.056701} {Phys. Rev. Lett. \textbf{132}, 056701 (2024)}.
\end{enumerate}

\end{small}

\end{document}